\documentclass[amsmath,aps,showpacs,a4paper,10pt]{revtex4}

 \usepackage{epsf}
 \usepackage{graphicx}    

\begin{document}

 \newcommand{\be}[1]{\begin{equation}\label{#1}}
 \newcommand{\ee}{\end{equation}}
 \newcommand{\bea}{\begin{eqnarray}}
 \newcommand{\eea}{\end{eqnarray}}
 \def\disp{\displaystyle}

 \def\gsim{ \lower .75ex \hbox{$\sim$} \llap{\raise .27ex \hbox{$>$}} }
 \def\lsim{ \lower .75ex \hbox{$\sim$} \llap{\raise .27ex \hbox{$<$}} }

 \begin{titlepage}

 \begin{flushright}
 arXiv:1106.0102
 \end{flushright}

 \title{\Large \bf $f(T)$ Theories and Varying Fine Structure
 Constant}

 \author{Hao~Wei\,}
 \email[\,email address:\ ]{haowei@bit.edu.cn}
 \affiliation{School of Physics, Beijing Institute
 of Technology, Beijing 100081, China}

 \author{Xiao-Peng~Ma\,}
 \affiliation{School of Physics, Beijing Institute
 of Technology, Beijing 100081, China}

 \author{Hao-Yu~Qi}
 \affiliation{School of Physics, Beijing Institute
 of Technology, Beijing 100081, China}

 \begin{abstract}\vspace{1cm}
 \centerline{\bf ABSTRACT}\vspace{2mm}
 In analogy to $f(R)$ theory, recently a new modified gravity
 theory, namely the so-called $f(T)$ theory, has been proposed
 to drive the current accelerated expansion without invoking
 dark energy. In the present work, by extending Bisabr's
 idea, we try to constrain $f(T)$ theories with the varying
 fine structure ``constant'', $\alpha\equiv e^2/\hbar c$. We
 find that the constraints on $f(T)$ theories from the
 observational $\Delta\alpha/\alpha$ data are very severe. In
 fact, they make $f(T)$ theories almost indistinguishable from
 $\Lambda$CDM model.
 \end{abstract}

 \pacs{04.50.Kd, 06.20.Jr, 98.80.-k, 95.36.+x}

 \maketitle

 \end{titlepage}

 \renewcommand{\baselinestretch}{1.1}


\section{Introduction}\label{sec1}

The current accelerated expansion of our universe~\cite{r1}
 has been one of the most active fields in modern cosmology
 since its discovery in 1998. This mysterious phenomenon
 could be due to an unknown energy component (dark energy)
 or a modification to general relativity (modified gravity)
 \cite{r1,r2,r3,r4}. The well-known modified gravity theories
 are, for examples, $f(R)$ theory, scalar-tensor theory
 (including Brans-Dicke theory), braneworld scenarios (such
 as DGP, RSI and RSII), $f(\cal G)$ theory ($\cal G$ is the
 Gauss-Bonett term), Horava-Lifshitz theory, MOND and TeVeS
 theories. We refer to e.g.~\cite{r1,r2,r3,r4,r50} for some
 reviews.

Recently, a new modified gravity theory, namely the so-called
 $f(T)$ theory, attracted much attention in the community,
 where $T$ is the torsion scalar. It is a generalized version
 of the so-called teleparallel gravity originally proposed by
 Einstein~\cite{r5,r6}. In teleparallel gravity,
 the Weitzenb\"ock connection is used, rather than the
 Levi-Civita connection which is used in general relativity.
 Following~\cite{r7,r8}, here we briefly review the key
 ingredients of teleparallel gravity and $f(T)$ theory.
 We consider a spatially flat Friedmann-Robertson-Walker (FRW)
 universe whose spacetime is described by
 \be{eq1}
 ds^2=-dt^2+a^2(t)d{\bf x}^2,
 \ee
 where $a$ is the scale factor. The orthonormal tetrad
 components $e_i(x^\mu)$ relate to the metric through
 \be{eq2}
 g_{\mu\nu}=\eta_{ij}e_\mu^i e_\nu^j\,,
 \ee
 where Latin $i$, $j$ are indices running over 0, 1, 2, 3 for
 the tangent space of the manifold, and Greek $\mu$,~$\nu$ are
 the coordinate indices on the manifold, also running over 0,
 1, 2, 3. In teleparallel gravity, the gravitational action is
 \be{eq3}
 {\cal S}_T=\frac{1}{2\kappa^2}\int d^4 x\,|e|\,T\,,
 \ee
 where $\kappa^2\equiv 8\pi G$, and
 $|e|={\rm det}\,(e_\mu^i)=\sqrt{-g}\,$. The torsion scalar $T$
 is given by
 \be{eq4}
 T\equiv{S_\rho}^{\mu\nu}\,{T^\rho}_{\mu\nu}\,,
 \ee
 where
 \bea
 {T^\rho}_{\mu\nu} &\equiv &-e^\rho_i\left(\partial_\mu e^i_\nu
 -\partial_\nu e^i_\mu\right)\,,\label{eq5}\\
 {K^{\mu\nu}}_\rho &\equiv &-\frac{1}{2}\left({T^{\mu\nu}}_\rho
 -{T^{\nu\mu}}_\rho-{T_\rho}^{\mu\nu}\right)\,,\label{eq6}\\
 {S_\rho}^{\mu\nu} &\equiv &\frac{1}{2}\left({K^{\mu\nu}}_\rho
 +\delta^\mu_\rho {T^{\theta\nu}}_\theta-
 \delta^\nu_\rho {T^{\theta\mu}}_\theta\right)\,.\label{eq7}
 \eea
 For a spatially flat FRW universe, from Eqs.~(\ref{eq4})
 and~(\ref{eq1}), one has
 \be{eq8}
 T=-6H^2,
 \ee
 where $H\equiv\dot{a}/a$ is the Hubble parameter (a dot
 denotes the derivative with respect to cosmic time $t$). So,
 one can use $T$ and $H$ interchangeably. In analogy to $f(R)$
 theory, one can replace $T$ in the gravitational
 action~(\ref{eq3}) by a function $f(T)$ (see
 however~\cite{r9}). In $f(T)$ theory, the modified Friedmann
 equation and Raychaudhuri equation are given by~\cite{r7,r8}
 \bea
 &&12H^2 f_T+f=16\pi G\rho\,,\label{eq9}\\
 &&48H^2 f_{TT}\dot{H}-f_T\left(12H^2+4\dot{H}\right)-f
 =16\pi Gp\,,\label{eq10}
 \eea
 where $f_T\equiv\partial f/\partial T$, and $\rho$, $p$ are
 the total energy density and pressure, respectively. In an
 universe with only dust matter, $p=p_m=0$ and $\rho=\rho_m$.
 From Eqs.~(\ref{eq9}) and~(\ref{eq10}), one can find that
 the effective dark energy density and pressure from torsion
 are given by~\cite{r7,r8,r10}
 \bea
 &&\rho_{de}=\frac{1}{16\pi G}\left(6H^2-f-
 12H^2 f_T\right)\,,\label{eq11}\\
 &&p_{de}=-\rho_{de}-\frac{1}{4\pi G}\left(12H^2 f_{TT}
 -f_T+1\right)\dot{H}\,.\label{eq12}
 \eea
 Obviously, if $f(T)=T+const.$, $f(T)$ theory reduces to the
 well-known $\Lambda$CDM model.

\newpage  

In fact, $f(T)$ theory was firstly used to drive inflation by
 Ferraro and Fiorini~\cite{r11,r12}. Later, Bengochea
 and Ferraro~\cite{r7}, as well as Linder~\cite{r8}, proposed
 to use $f(T)$ theory to drive the current accelerated
 expansion without invoking dark energy. Soon, many works
 followed. For examples, Myrzakulov~\cite{r13} and
 Yang~\cite{r14} proposed some new $f(T)$ forms;
 Bengochea~\cite{r15}, Wu and Yu~\cite{r16} considered the
 cosmological constraints on $f(T)$ theories by using the
 latest observational data; Wu and Yu~\cite{r17}
 also considered the dynamical behavior of $f(T)$ theory;
 Dent~{\it et al.}~\cite{r18}, Zheng and Huang~\cite{r19}
 considered the cosmological perturbations and growth factor
 in $f(T)$ theories; Wu and Yu~\cite{r20}, Bamba
 and Geng~\cite{r21} discussed the equation-of-state parameter
 (EoS) crossing the phantom divide in $f(T)$ theories;
 Zhang~{\it et al.}~\cite{r22} discussed the dynamical analysis
 of $f(T)$ theories; Li, Sotiriou and Barrow~\cite{r23}
 considered the large-scale structure and local Lorentz
 invariance in $f(T)$ theory; Deliduman and Yapiskan~\cite{r24}
 discussed the relativistic neutron star in $f(T)$ theory;
 Cai~{\it et al.}~\cite{r25} considered the matter bounce in
 $f(T)$ theory; Wang~\cite{r26} discussed the static solutions
 with spherical symmetry in $f(T)$ theories. We further refer
 to e.g.~\cite{r27} for some relevant works.

In the literature, the observational constraints on $f(T)$
 theories~\cite{r15,r16} were obtained mainly by using the
 cosmological data, such as type Ia supernovae (SNIa), baryon
 acoustic oscillation (BAO), and cosmic microwave background
 (CMB). In the present work, we instead try to constrain
 $f(T)$ theories with the varying fine structure
 ``constant'', $\alpha\equiv e^2/\hbar c$. In Sec.~\ref{sec2},
 we briefly review the observational constraints on the
 temporal variation of the fine structure ``constant''
 $\alpha$. In Sec.~\ref{sec3}, we briefly introduce the idea
 to constrain $f(R)$ theories with the varying $\alpha$, which
 was proposed by Bisabr~\cite{r28}. In Sec.~\ref{sec4}, we
 extend Bisabr's idea to $f(T)$ theories, and consider the
 corresponding constraints from the temporal variation of the
 fine structure ``constant''. Finally, some brief
 concluding remarks are given in Sec.~\ref{sec5}.


 \begin{table}[htbp]
 \begin{center}
 \vspace{3mm}  
 \begin{tabular}{cccc} \hline\hline
 ~~$|\Delta\alpha/\alpha|$~~~ & ~~~Redshift~~~
 & ~~~Observation~~~ & \hspace{6mm} Ref.\hspace{6mm} \\ \hline\\[-4mm]
 $\lsim\, 10^{-2}$ & $10^{10}-10^8$ & BBN & \cite{r33,r34}\\
 $< 10^{-2}$ & $10^3$ & CMB & \cite{r34}\\
 $\lsim\, 10^{-6}$ & $3-0.4$ & quasars & \cite{r30,r31,r35,r36}\\
 $\lsim\, 10^{-7}$ & $0.45$ & meteorite & \cite{r37}\\
 $\lsim\, 10^{-7}$ & $0.14$ & Oklo & \cite{r38}\\[0.5mm]
 \hline\hline
 \end{tabular}
 \end{center}
 \caption{\label{tab1} The observational constraints on
 $\Delta\alpha/\alpha$.}
 \end{table}


\vspace{-6mm}  


\section{Observational constraints on the temporal variation of
 the~fine~structure ``constant''}\label{sec2}

Motivated by the well-known large number hypothesis of
 Dirac proposed in 1937~\cite{r29}, the varying fundamental
 ``constants'' remain as one of the unfading subjects for
 decades. Among the fundamental ``constants'', the most
 observationally sensitive one is the electromagnetic fine
 structure ``constant'', $\alpha\equiv e^2/\hbar c$. Since
 about 12 years ago, this subject attracted many attentions,
 mainly due to the first observational evidence from the
 quasar absorption spectra that the fine structure
 ``constant'' might change with cosmological
 time~\cite{r30,r31}.

Subsequently, many authors obtained various observational
 constraints on the temporal variation of the fine structure
 ``constant'' $\alpha$. In the literature, it is convenient
 to introduce a quantity
 $\Delta\alpha/\alpha\equiv (\alpha-\alpha_0)/\alpha_0$,
 where the subscript ``0'' indicates the present value of the
 corresponding quantity. Obviously, $\Delta\alpha/\alpha$ is
 time-dependent. A brief summary of the observational
 constraints on $\Delta\alpha/\alpha$ can be found in
 e.g.~\cite{r32}. The most ancient constraint comes from the
 Big Bang Nucleosynthesis (BBN)~\cite{r33,r34}, namely,
 $|\Delta\alpha/\alpha|\,\lsim\, 10^{-2}$, in the redshift
 range $z=10^{10}-10^8$. The next constraint comes from the
 power spectrum of anisotropy in the cosmic microwave
 background (CMB)~\cite{r34}, i.e.,
 $|\Delta\alpha/\alpha| < 10^{-2}$, for redshift
 $z\simeq 10^3$. In the medium redshift range, the constraint
 comes from the absorption spectra of distant
 quasars~\cite{r30,r31,r35,r36}. Since the results in the
 literature are controversial, it is better to consider the
 conservative constraint
 $|\Delta\alpha/\alpha|\,\lsim\, 10^{-6}$~\cite{r32}, in the
 redshift range $z=3-0.4$. From the radioactive life-time of
 $^{187}$Re derived from meteoritic studies~\cite{r37}, the
 constraint is given by $|\Delta\alpha/\alpha|\,\lsim\, 10^{-7}$
 for redshift $z=0.45$. Finally, from the Oklo natural nuclear
 reactor~\cite{r38}, it is found that
 $|\Delta\alpha/\alpha|\,\lsim\, 10^{-7}$ for redshift $z=0.14$.
 For convenience, we summarize the above constraints in
 Table~\ref{tab1}, which will be used in the followings.


\section{The idea to constrain $f(R)$ theories
 with varying alpha}\label{sec3}

\subsection{Varying alpha driven by a general
 scalar field}\label{sec3a}

Noting that $\alpha=e^2/\hbar c$, a varying $\alpha$ might be
 due to a varying speed of light $c$~\cite{r39,r40,r41}, while
 Lorentz invariance is broken. The other possibility for
 a varying $\alpha$ is due to a varying electron charge $e$.
 In 1982, Bekenstein proposed such a varying $\alpha$
 model~\cite{r42}, which preserves local gauge and Lorentz
 invariance, and is generally covariant. This model has been
 revived and generalized after the first observational
 evidence of varying $\alpha$ from the quasar absorption
 spectra~\cite{r30,r31}. This is a dilaton theory with coupling
 to the electromagnetic $F^2$ part of the Lagrangian, but not
 to the other gauge fields. Later, the Bekenstein-type varying
 $\alpha$ model has been generalized by replacing the dilaton
 with a cosmological scalar field. Further, the coupling
 between the scalar field and the electromagnetic field could
 also be generalized. In fact, the varying $\alpha$ models
 driven by quintessence have been extensively investigated in
 the literature (see e.g.~\cite{r32,r43,r44,r45,r46}). The
 varying $\alpha$ driven by phantom has been considered in the
 BSBM model~\cite{r47} while its model parameter $\omega$ is
 negative. The special case of varying $\alpha$ driven by
 $k$-essence whose Lagrangian ${\cal L}(X,\phi)=X^n-V(\phi)$
 has been considered in e.g.~\cite{r43}. The varying $\alpha$
 driven by Dirac-Born-Infeld scalar field has also been
 discussed in~\cite{r48}.

Following~\cite{r45,r32,r43}, the relevant action in Einstein
 frame is generally given by
 \be{eq13}
 {\cal S}={\cal S}_g+\int d^4 x\,\sqrt{-g}\,{\cal L}_\phi-
 \frac{1}{4}\int d^4 x\,\sqrt{-g}\,B_F(\phi)\,F_{\mu\nu}F^{\mu\nu}+
 {\cal S}_m\,,
 \ee
 where ${\cal S}_g$ is the gravitational action in Einstein
 frame; $F_{\mu\nu}$ are the components of the electromagnetic
 field tensor; ${\cal S}_m$ is the action of other matters;
 ${\cal L}_\phi$ is the Lagrangian of the scalar field $\phi$.
 Noting that $B_F$ takes the place of $e^{-2}$ in
 Eq.~(\ref{eq13}) actually~\cite{r44,r49}, one can easily see
 that the effective fine structure ``constant''
 $\alpha=e^2/\hbar c$ is given by~\cite{r32,r43}
 \be{eq14}
 \alpha=\frac{\alpha_0}{B_F(\phi({\bf x},t))}\,.
 \ee
 Thus, we find that
 \be{eq15}
 \frac{\Delta\alpha}{\alpha}\equiv\frac{\alpha
 -\alpha_0}{\alpha_0}=\frac{1}{B_F(\phi)}-1\,.
 \ee
 It is worth noting that the present value (at redshift $z=0$)
 of the coupling $B_F$ should be $1$ by definition. If
 $B_F(z=0)=B_{F0}\not=1$, we can normalize it through rescaling
 the electromagnetic field, namely
 \be{eq16}
 F_{\mu\nu}\to\sqrt{B_{F0}}\,F_{\mu\nu}\,,~~~~~~~{\rm while}
 ~~~~~~~B_F\to\frac{B_F}{\,B_{F0}}\,.
 \ee
 Thus, we have
 \be{eq17}
 \frac{\Delta\alpha}{\alpha}=\frac{B_{F0}}{B_F}-1\,.
 \ee

\subsection{Varying alpha in $f(R)$ theories}\label{sec3b}

Following Bisabr's idea~\cite{r28}, here we briefly show why
 the fine structure ``constant'' should be varying in $f(R)$
 theories. As is well known, in Jordan frame the action of
 $f(R)$ theories with matters (including electromagnetic
 fields here) reads (see e.g.~\cite{r2,r3,r4})
 \be{eq18}
 {\cal S}=\frac{1}{2\kappa^2}\int d^4x\sqrt{-g}\,f(R)+
 \int d^4x\,{\cal L}_M\left(g_{\mu\nu},\Psi_M\right),
 \ee
 where $R$ is the Ricci scalar, and ${\cal L}_M$ is the matter
 Lagrangian depending on $g_{\mu\nu}$ and matter fields
 $\Psi_M$ (including electromagnetic fields here). It is well
 known that $f(R)$ theory can be equivalent to scalar-tensor
 theory~\cite{r2,r3,r4}. Applying the conformal transformation
 \be{eq19}
 \tilde{g}_{\mu\nu}=\Omega^2 g_{\mu\nu}\,,~~~~~~~
 \Omega^2={\cal F}\equiv f_R=\frac{\partial f}{\partial R}\,,
 \ee
 and introducing a new scalar field $\phi$ defined by
 \be{eq20}
 \kappa\phi\equiv\sqrt{3/2}\,\ln{\cal F}\,,
 \ee
 one can rewrite the action (\ref{eq18}) to the one in Einstein
 frame~\cite{r2,r3,r4}, namely
 \be{eq21}
 \tilde{\cal S}=\int d^4x\sqrt{-\tilde{g}}\left[
 \frac{\tilde{R}}{2\kappa^2}-
 \frac{1}{2}\tilde{g}^{\mu\nu}\partial_\mu\phi\partial_\nu\phi
 -V(\phi)\right]+\int d^4x\,{\cal L}_M\left(
 {\cal F}^{-1}(\phi)\,\tilde{g}_{\mu\nu},\Psi_M\right)\,,
 \ee
 where a tilde denotes quantities in Einstein frame, and the
 potential of scalar field is given by
 \be{eq22}
 V(\phi)=\frac{{\cal F}R-f}{2\kappa^2{\cal F}^2}\,.
 \ee
 From Eq.~(\ref{eq21}), it is easy to see that the matter
 fields (including electromagnetic fields here)
 are {\em inevitably} coupled with the scalar field $\phi$ in
 Einstein frame. So, $\alpha$ should be varying. Noting that
 $\sqrt{-g}=\Omega^{-4}\sqrt{-\tilde{g}}$, one can easily
 find that the coupling in Eq.~(\ref{eq13}) is given
 by~\cite{r28}
 \be{eq23}
 B_F={\cal F}^{-2}=f_R^{-2}.
 \ee
 From Eq.~(\ref{eq15}), the variation of the fine structure
 ``constant'' can be described by~\cite{r28}
 \be{eq24}
 \frac{\Delta\alpha}{\alpha}=f_R^2-1\,.
 \ee
 As mentioned above, if $f_R(z=0)=f_{R0}\not=1$, we can
 normalize it through Eq.~(\ref{eq16}), and then
 \be{eq25}
 \frac{\Delta\alpha}{\alpha}=\left(\frac{f_R}{f_{R0}}\right)^2-1\,.
 \ee
 In~\cite{r28}, Bisabr discussed the observational constraints
 on $f(R)$ theories with the temporal variation of the fine
 structure ``constant'', and found that the corresponding
 constraints are fairly tight. We refer to the original
 paper~\cite{r28} for details.


\section{$f(T)$ theories and varying alpha}\label{sec4}

As mentioned in Sec.~\ref{sec1}, $f(T)$ theories are proposed
 in analogy to $f(R)$ theories. So, we extend Bisabr's
 idea~\cite{r28} to constrain $f(T)$ theories also with the
 temporal variation of the fine structure ``constant''.

\subsection{Varying alpha in a general $f(T)$ theory}\label{sec4a}

As mentioned in~\cite{r8,r10}, $f(T)$ theory can also be
 equivalent to scalar-tensor (torsion) theory. In Jordan frame,
 the relevant action with matters (including electromagnetic
 fields here) reads
 \be{eq26}
 {\cal S}=\frac{1}{2\kappa^2}\int d^4 x\,|e|\,f(T)
 +\int d^4x\,{\cal L}_M\left(e_\mu^i,\Psi_M\right).
 \ee
 Similarly, applying the conformal transformation
 \be{eq27}
 \tilde{g}_{\mu\nu}=\Omega^2 g_{\mu\nu}~\leftrightarrow~\,
 \tilde{e}_\mu^i=\Omega\,e_\mu^i\,,~~~~~~~
 \Omega^2={\cal F}\equiv f_T\,,
 \ee
 one can also rewrite the action (\ref{eq26}) to the one in
 Einstein frame with a new scalar field $\phi$~\cite{r8,r10}.
 Similar to the case of $f(R)$ theories, the matter action
 in Einstein frame is given by~\cite{r10}
 \be{eq28}
 \tilde{\cal S}_M=\int d^4x\,{\cal L}_M\left(
 {\cal F}^{-1/2}(\phi)\,\tilde{e}_\mu^i,\Psi_M\right).
 \ee
 Again, the matter fields (including electromagnetic fields
 here) are {\em inevitably} coupled with the scalar field
 $\phi$ in Einstein frame. So, $\alpha$ should be varying.
 Noting that $|e|=\Omega^{-4}|\tilde{e}|$, one can similarly
 find that
 \be{eq29}
 B_F={\cal F}^{-2}=f_T^{-2}.
 \ee
 From Eq.~(\ref{eq15}), the variation of the fine structure
 ``constant'' can be described by
 \be{eq30}
 \frac{\Delta\alpha}{\alpha}=f_T^2-1\,.
 \ee
 As mentioned above, if $f_T(z=0)=f_{T0}\not=1$, we can
 normalize it through Eq.~(\ref{eq16}), and then
 \be{eq31}
 \frac{\Delta\alpha}{\alpha}=\left(\frac{f_T}{f_{T0}}\right)^2-1\,.
 \ee

In the followings, we will consider two concrete $f(T)$
 theories, namely, $f(T)=T+\mu(-T)^n$ and
 $f(T)=T-\mu T\left(1-e^{\beta T_0/T}\right)$, which are the
 most popular $f(T)$ theories discussed extensively in
 the literature (see e.g.~\cite{r7,r8,r16,r17}).


 \begin{center}
 \begin{figure}[htbp]
 \centering
 \includegraphics[width=0.98\textwidth]{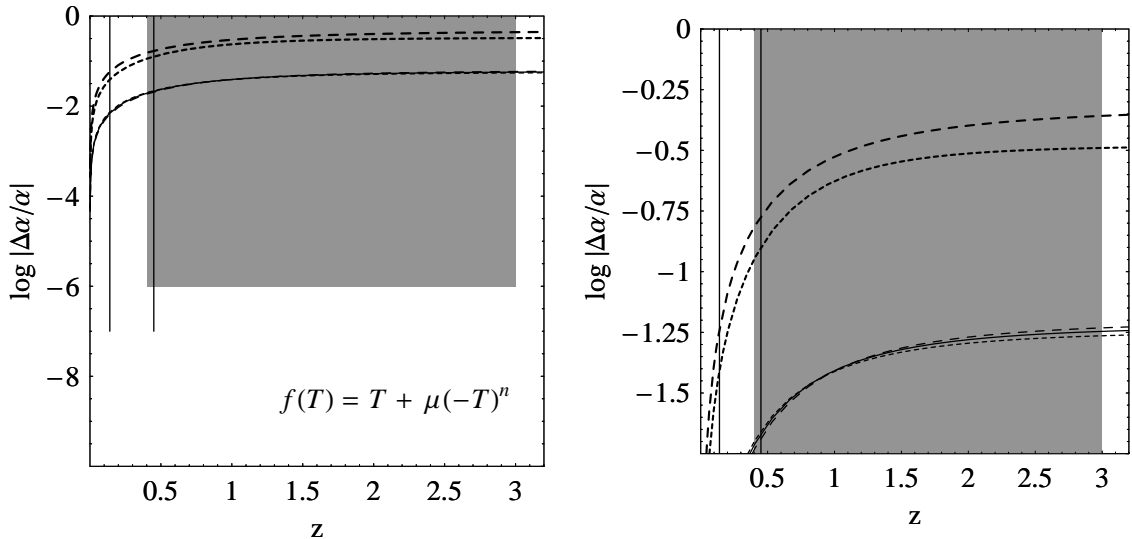}
 \caption{\label{fig1} $\log |\Delta\alpha/\alpha|$ as a
 function of redshift $z$ for $f(T)=T+\mu(-T)^n$ with
 $\Omega_{m0}=0.272$ and $n=0.04$ (solid curve),
 $\Omega_{m0}=0.272$ and $n=0.26$ (thick long-dashed curve),
 $\Omega_{m0}=0.272$ and $n=-0.29$ (thick short-dashed curve),
 $\Omega_{m0}=0.240$ and $n=0.04$ (thin long-dashed curve),
 $\Omega_{m0}=0.308$ and $n=0.04$ (thin short-dashed curve).
 Right panel is an enlarged part of left panel. Only the
 curves not overlapping the gray areas are phenomenologically
 viable. See text for details.}
 \end{figure}
 \end{center}


\vspace{-12mm}  


 \begin{center}
 \begin{figure}[tbp]
 \centering
 \includegraphics[width=0.45\textwidth]{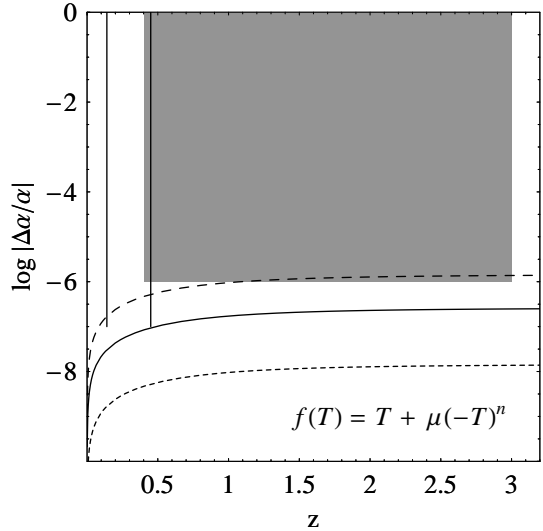}
 \caption{\label{fig2} $\log |\Delta\alpha/\alpha|$ as a
 function of redshift $z$ for $f(T)=T+\mu(-T)^n$ with a
 fixed $\Omega_{m0}=0.272$, and
 $n=\pm 1.8\times 10^{-7}$ (solid curve),
 $n=\pm 10^{-6}$ (long-dashed curve),
 $n=\pm 10^{-8}$ (short-dashed curve).
 Only the curves not overlapping the gray areas
 are phenomenologically viable. See text for details.}
 \end{figure}
 \end{center}


\vspace{-12mm}  

\subsection{$f(T)=T+\mu(-T)^n$}\label{sec4b}

At first, we consider the case of $f(T)=T+\mu(-T)^n$, where
 $\mu$ and $n$ are both constants. This is the simplest model,
 and has been considered in most papers on $f(T)$ theory.
 Obviously, if $n=0$, it reduces to $\Lambda$CDM model.
 Substituting it into the modified Friedmann
 equation~(\ref{eq9}), one can easily find that $\mu$ is not
 an independent model parameter, namely~\cite{r8,r16}
 \be{eq32}
 \mu=\frac{1-\Omega_{m0}}{2n-1}\left(6H_0^2\right)^{1-n}=
 \frac{1-\Omega_{m0}}{2n-1}\left(-T_0\right)^{1-n},
 \ee
 where $\Omega_{m0}\equiv 8\pi G\rho_{m0}/(3H_0^2)$ is the
 present fractional energy density of dust matter. So, we have
 \be{eq33}
 f(T)=T+\frac{1-\Omega_{m0}}{2n-1}(-T_0)\left(
 \frac{T}{T_0}\right)^n\,,
 \ee
 and then
 \be{eq34}
 f_T=1+\frac{n(1-\Omega_{m0})}{1-2n}E^{2(n-1)}\,,~~~~~~~
 f_{T0}=1+\frac{n(1-\Omega_{m0})}{1-2n}\,,
 \ee
 where $E^2=T/T_0=H^2/H_0^2$. Substituting Eq.~(\ref{eq34})
 into Eq.~(\ref{eq31}), one can finally obtain the explicit
 expression of $\Delta\alpha/\alpha$. In order to compare
 it with the observational data, we need to know $E(z)$ as
 a function of redshift $z$. Substituting $f(T)=T+\mu(-T)^n$
 and Eq.~(\ref{eq32}) into the modified Friedmann
 equation~(\ref{eq9}), we find that
 \be{eq35}
 E^2=\Omega_{m0}(1+z)^3+(1-\Omega_{m0})E^{2n}\,.
 \ee
 Obviously, if $n=0$, it reduces to the one of $\Lambda$CDM
 model. If $\Omega_{m0}$ and $n$ are given, we can numerically
 solve Eq.~(\ref{eq35}) and obtain $E^2(z)$ as a function of
 redshift $z$. Thus, $\Delta\alpha/\alpha$ is on hand.

Next, we compare $\Delta\alpha/\alpha$ with the observational
 data. In fact, as shown in e.g.~\cite{r32,r48}, the
 constraints from the first two rows (at very high redshift)
 in Table~\ref{tab1} are very weak. Therefore, we only consider
 the last three rows (at low redshift) in Table~\ref{tab1}
 (and hence the radiation can be safely ignored). Note that
 in~\cite{r16}, this $f(T)=T+\mu(-T)^n$ model has been
 constrained by using the latest cosmological data, i.e., 557
 Union2 SNIa dataset, BAO, and shift parameter from WMAP7.
 The corresponding $2\sigma$ results are given by~\cite{r16}
 \be{eq36}
 \Omega_{m0}=0.272^{+0.036}_{-0.032}\,,~~~~~~~
 n=0.04^{+0.22}_{-0.33}\,.
 \ee
 At first, we try to see whether $\Delta\alpha/\alpha$ with
 the best-fit parameters of~\cite{r16} and the corresponding
 $2\sigma$ edge can simultaneously satisfy the observational
 constraints in Table~\ref{tab1}. In Fig.~\ref{fig1}, we plot
 $\log |\Delta\alpha/\alpha|$ as a
 function of redshift $z$ for $f(T)=T+\mu(-T)^n$ with
 $\Omega_{m0}=0.272$ and $n=0.04$ (solid curve),
 $\Omega_{m0}=0.272$ and $n=0.26$ (thick long-dashed curve),
 $\Omega_{m0}=0.272$ and $n=-0.29$ (thick short-dashed curve),
 $\Omega_{m0}=0.240$ and $n=0.04$ (thin long-dashed curve),
 $\Omega_{m0}=0.308$ and $n=0.04$ (thin short-dashed curve),
 where log indicates the logarithm to base $10$. Obviously,
 one can see that $\Delta\alpha/\alpha$ with the
 best-fit parameters of~\cite{r16} and the corresponding
 $2\sigma$ edge {\em cannot} satisfy the observational
 constraints in Table~\ref{tab1}. In addition,
 from Fig.~\ref{fig1}, we find that the influence from $n$
 to $\Delta\alpha/\alpha$ is significantly larger than the
 one from $\Omega_{m0}$. Thus, fixing $\Omega_{m0}=0.272$,
 we try various $n$ to find in which cases all the
 observational constraints in Table~\ref{tab1} could be
 simultaneously satisfied. From Fig.~\ref{fig2}, it is easy
 to see that they can be all respected only for
 \be{eq37}
 |n|\leq 1.8\times 10^{-7}\,.
 \ee
 This is the constraint on $f(T)=T+\mu(-T)^n$ theory from the
 observational $\Delta\alpha/\alpha$ data. It is a very severe
 constraint in fact. Noting that $f(T)=T+\mu(-T)^n\to T+const.$
 when $n\to 0$, this $f(T)$ theory becomes almost
 indistinguishable from $\Lambda$CDM model.


 \begin{center}
 \begin{figure}[htbp]
 \centering
 \includegraphics[width=0.98\textwidth]{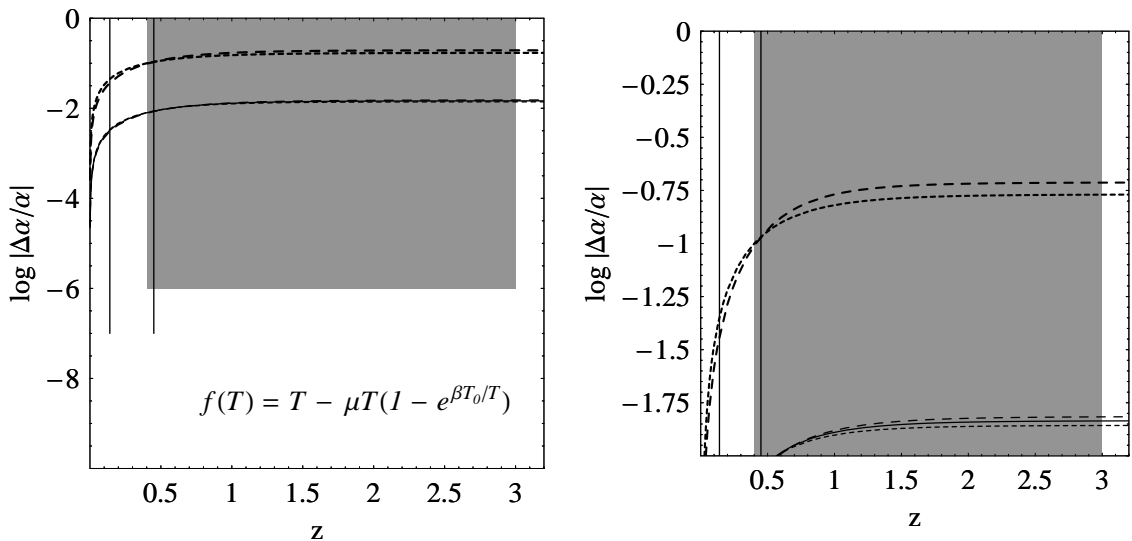}
 \caption{\label{fig3} $\log |\Delta\alpha/\alpha|$ as a
 function of redshift $z$ for
 $f(T)=T-\mu T\left(1-e^{\beta T_0/T}\right)$ with
 $\Omega_{m0}=0.272$ and $\beta=-0.02$ (solid curve),
 $\Omega_{m0}=0.272$ and $\beta=0.29$ (thick long-dashed curve),
 $\Omega_{m0}=0.272$ and $\beta=-0.22$ (thick short-dashed curve),
 $\Omega_{m0}=0.238$ and $\beta=-0.02$ (thin long-dashed curve),
 $\Omega_{m0}=0.308$ and $\beta=-0.02$ (thin short-dashed curve).
 Right panel is an enlarged part of left panel. Only the
 curves not overlapping the gray areas are phenomenologically
 viable. See text for details.}
 \end{figure}
 \end{center}


\vspace{-11mm}  


 \begin{center}
 \begin{figure}[tbp]
 \centering
 \includegraphics[width=0.45\textwidth]{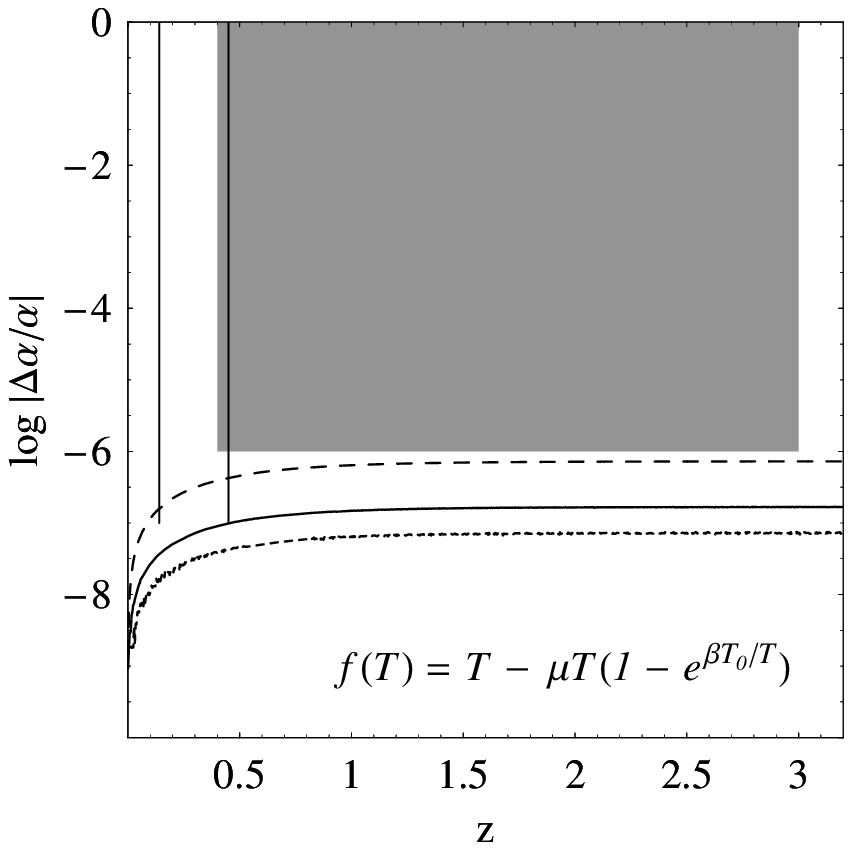}
 \caption{\label{fig4} $\log |\Delta\alpha/\alpha|$ as a
 function of redshift $z$ for
 $f(T)=T-\mu T\left(1-e^{\beta T_0/T}\right)$
 with a fixed $\Omega_{m0}=0.272$, and
 $\beta=\pm 2.3\times 10^{-7}$ (solid curve),
 $\beta=\pm 10^{-6}$ (long-dashed curve),
 $\beta=\pm 10^{-7}$ (short-dashed curve).
 Only the curves not overlapping the gray areas
 are phenomenologically viable. See text for details.}
 \end{figure}
 \end{center}


\vspace{-11mm}  

\subsection{$f(T)=T-\mu T\left(1-e^{\beta T_0/T}\right)$}\label{sec4c}

Here, we consider the case of
 $f(T)=T-\mu T\left(1-e^{\beta T_0/T}\right)$, where $\mu$
 and $\beta$ are both constants.
 Obviously, $f(T)\to T+\mu\beta T_0=T+const.$ when
 $\beta\to 0$, it reduces to $\Lambda$CDM model. Substituting
 it into the modified Friedmann equation~(\ref{eq9}), one can
 easily find that $\mu$ is not an independent
 model parameter~\cite{r8,r16}, i.e.,
 \be{eq38}
 \mu=\frac{1-\Omega_{m0}}{1-\left(1-2\beta\right)e^\beta}\,.
 \ee
 On the other hand, it is easy to obtain
 \be{eq39}
 f_T=1-\mu+\mu\left(1-\beta/E^2\right)e^{\beta/E^2}\,,~~~~~~~
 f_{T0}=1-\mu+\mu\left(1-\beta\right)e^\beta\,,
 \ee
 where $E^2=T/T_0=H^2/H_0^2$. Substituting Eq.~(\ref{eq39})
 into Eq.~(\ref{eq31}), one can finally obtain the explicit
 expression of $\Delta\alpha/\alpha$. In order to compare
 it with the observational data, we need to know $E(z)$ as
 a function of redshift $z$. Substituting
 $f(T)=T-\mu T\left(1-e^{\beta T_0/T}\right)$ into the modified
 Friedmann equation~(\ref{eq9}), we find that
 \be{eq40}
 E^2=\Omega_{m0}(1+z)^3+\mu E^2\left[
 1-e^{\beta/E^2}+2\left(\frac{\beta}{E^2}\right)e^{\beta/E^2}
 \right]\,.
 \ee
 If $\beta\to 0$, we have $\mu\beta\to 1-\Omega_{m0}$ from
 Eq.~(\ref{eq38}), and hence Eq.~(\ref{eq40}) reduces to the
 one of $\Lambda$CDM model. If $\Omega_{m0}$ and $\beta$ are
 given, we can numerically solve Eq.~(\ref{eq40}) and obtain
 $E^2(z)$ as a function of redshift $z$. Thus,
 $\Delta\alpha/\alpha$ is on hand.

In~\cite{r16}, this
 $f(T)=T-\mu T\left(1-e^{\beta T_0/T}\right)$ model has also
 been constrained by using the latest cosmological data, i.e.,
 557 Union2 SNIa dataset, BAO, and shift parameter from WMAP7.
 The corresponding $2\sigma$ results are given by~\cite{r16}
 \be{eq41}
 \Omega_{m0}=0.272^{+0.036}_{-0.034}\,,~~~~~~~
 \beta=-0.02^{+0.31}_{-0.20}\,.
 \ee
 Again, we try to see whether $\Delta\alpha/\alpha$ with
 the best-fit parameters of~\cite{r16} and the corresponding
 $2\sigma$ edge can simultaneously satisfy the observational
 constraints in Table~\ref{tab1}. In Fig.~\ref{fig3}, we plot
 $\log |\Delta\alpha/\alpha|$ as a
 function of redshift $z$ for
 $f(T)=T-\mu T\left(1-e^{\beta T_0/T}\right)$ with
 $\Omega_{m0}=0.272$ and $\beta=-0.02$ (solid curve),
 $\Omega_{m0}=0.272$ and $\beta=0.29$ (thick long-dashed curve),
 $\Omega_{m0}=0.272$ and $\beta=-0.22$ (thick short-dashed curve),
 $\Omega_{m0}=0.238$ and $\beta=-0.02$ (thin long-dashed curve),
 $\Omega_{m0}=0.308$ and $\beta=-0.02$ (thin short-dashed curve).
 Obviously, one can see that $\Delta\alpha/\alpha$ with the
 best-fit parameters of~\cite{r16} and the corresponding
 $2\sigma$ edge {\em cannot} satisfy the observational
 constraints in Table~\ref{tab1}. In addition,
 from Fig.~\ref{fig3}, we find that the influence from $\beta$
 to $\Delta\alpha/\alpha$ is significantly larger than the one
 from $\Omega_{m0}$. Thus, fixing $\Omega_{m0}=0.272$, we try
 various $\beta$ to find in which cases all the observational
 constraints in Table~\ref{tab1} could be simultaneously
 satisfied. From Fig.~\ref{fig4}, it is easy to see that they
 can be all respected only for
 \be{eq42}
 |\beta|\leq 2.3\times 10^{-7}\,.
 \ee
 This is the constraint
 on $f(T)=T-\mu T\left(1-e^{\beta T_0/T}\right)$ theory from
 the observational $\Delta\alpha/\alpha$ data. It is a very
 severe constraint in fact. Noting
 that $f(T)=T-\mu T\left(1-e^{\beta T_0/T}\right)\to T+const.$
 when $\beta\to 0$, this $f(T)$ theory becomes almost
 indistinguishable from $\Lambda$CDM model.


\section{Concluding remarks}\label{sec5}

In analogy to $f(R)$ theory, recently $f(T)$ theory has
 been proposed to drive the current accelerated expansion
 without invoking dark energy. In the literature, the
 observational constraints on $f(T)$ theories were obtained
 mainly by using the cosmological data, such as type Ia
 supernovae (SNIa), baryon acoustic oscillation (BAO), and
 cosmic microwave background (CMB). In the present work, by
 extending Bisabr's idea~\cite{r28}, we instead try to
 constrain $f(T)$ theories with the varying fine structure
 ``constant'', $\alpha\equiv e^2/\hbar c$. We found that the
 constraints on $f(T)$ theories from the observational
 $\Delta\alpha/\alpha$ data are very severe. In fact, they make
 $f(T)$ theories almost indistinguishable from $\Lambda$CDM model.

Some remarks are in order. Firstly, in this work we only
 considered a spatially flat FRW universe. This is mainly motivated
 by the well-known inflation scenario and the very tight
 observational constraint on the spatial curvature term from WMAP7
 data~\cite{r51}, namely, $\Omega_k=-0.0057^{+0.0067}_{-0.0068}$.
 Obviously, when a non-vanishing spatial curvature term is allowed,
 the observational constraints on $f(T)$ theories from the
 varying fine structure ``constant'' should be relaxed, since
 the number of free model parameters is increased. However, we
 can expect that the situation of $f(T)$ theories are still not
 improved, due to the very narrow range of the constrained
 $\Omega_k$. For instance, even the constraints on $n$ or
 $\beta$ could be greatly relaxed from ${\cal O}(10^{-7})$ to
 ${\cal O}(10^{-4})$ (say), the corresponding $f(T)$ theories
 are still indistinguishable from $\Lambda$CDM model. Secondly,
 in fact the fine structure ``constant'' might be not only
 time-dependent but also space-dependent (see e.g.~\cite{r52}).
 Of course, the space-dependent fine structure ``constant'' is
 still in controversy. On the other hand, the space-dependent
 fine structure ``constant'' might invoke an inhomogeneous
 scalar field $\phi$. As is shown in Secs.~\ref{sec3b} and
 \ref{sec4a}, the $f(R)$ and $f(T)$ theories in a homogeneous
 and isotropic FRW universe could not lead to a space-dependent
 fine structure ``constant'', and hence they are not constrained by
 the possibly spatial variation of the fine structure ``constant''.
 So, in this work we only considered the time-dependent fine
 structure ``constant'' for simplicity. Thirdly, it is worth noting
 that $f(T)$ gravity does not generally preserve the local Lorentz
 invariance and any theory of $f(T)$ gravity is always built on
 a local frame which is chosen on a specific spacetime point.
 As consequences, it is not always applicable for the conformal
 transformation in $f(R)$ gravity to be used in $f(T)$ gravity,
 and it is quite unclear how to explicitly define an Einstein
 or a Jordan frame in $f(T)$ gravity. Fortunately, we could
 use these conceptions in the homogenous and isotropic FRW
 background which is quite a special case (we thank the referee
 for pointing out this issue). Finally, we note that the two
 concrete $f(T)$ theories considered in this work contain only
 a single free parameter, which are very simple cases. In fact,
 it is reminiscent of the case of $f(R)$ theory. In the
 beginning, the forms of $f(R)$ are also very simple, such as
 the types of $1/R$ or $R^n$~\cite{r1,r2,r3,r4}. Later, these
 simple $f(R)$ forms have been easily ruled out by the
 cosmological observations and the local gravity tests. After
 several years, the viable $f(R)$ forms which can satisfy all
 the cosmological observations and the local gravity tests,
 e.g. Hu-Sawicki~\cite{r53}, Starobinsky~\cite{r54}
 and Tsujikawa~\cite{r55}, have been toughly
 earned~\cite{r1,r2,r3,r4}. These three viable $f(R)$ forms are
 all complicated and delicate, and they contain two or more
 free parameters. Similarly, we expect that the viable
 $f(T)$ forms, which satisfy all the cosmological observations,
 the local gravity tests and the constraints from the varying
 fine structure ``constant'', could be constructed with tough
 efforts in the future. Of course, it is easy to anticipate
 that they are also complicated and delicate, and contain many
 free parameters. In addition, at that time it is also
 interesting to see what kind of dark energy models could be
 mimicked by the viable $f(T)$ theories.


\section*{ACKNOWLEDGEMENTS}
We thank the referee for quite useful comments and suggestions,
 which help us to improve this work. We are grateful to
 Professors Rong-Gen~Cai and Shuang~Nan~Zhang
 for helpful discussions. We also thank Minzi~Feng, as well as
 Puxun~Wu and Rong-Jia~Yang, for kind help and discussions.
 This work was supported in part by NSFC under Grant No.~10905005,
 and the Fundamental Research Fund of Beijing Institute of
 Technology.

\newpage  

\renewcommand{\baselinestretch}{1.15}



\begin{thebibliography}{99}
\bibitem{r9}
In the literature, one can instead write this function as $T+f(T)$,
 rather than $f(T)$ used here. One should be aware of the
 correspondence between these two formalisms.
\end{thebibliography}
\end{document}